\documentstyle[11pt]{article}

\advance \textheight by \topskip
\makeatletter
\@addtoreset{equation}{section}
  
\makeatother

\def\cc#1{\kern .7em\hfill #1 \hfill\kern .7em}  
\def\ZZ{\hbox{\it Z\hskip -4.pt Z}}

\def\RR{\hbox{\it I\hskip -2.pt R }}

\newcommand{\nc}{\newcommand}
\nc{\beq}{\begin{equation}}
\nc{\eeq}{\end{equation}}
\nc{\beqa}{\begin{eqnarray}}
\nc{\eeqa}{\end{eqnarray}}
\nc{\noi}{\noindent}

\begin{document}

\title{
\begin{flushright}
{\small USACH-FM-00/01}\\[-0.4cm]
{\small PM-00/01}\\[2cm]
\end{flushright}
{\bf Cubic root of Klein-Gordon equation}}

\author{{\sf
Mikhail S. Plyushchay}\thanks{e-mail: mplyushc@lauca.usach.cl}$\,\,$${}^{a,b}$
 and
{\sf Michel~Rausch de Traubenberg}\thanks{e-mail:
rausch@lpm.univ-montp2.fr}$\,\,$$^{c,d}$\\
\\
{\small ${}^{a}${\it Departamento de F\'{\i}sica, 
Universidad de Santiago de Chile,
Casilla 307, Santiago 2, Chile}}\\
{\small ${}^{b}${\it Institute for High Energy Physics, Protvino,
Russia}}\\
{\small ${}^{c}${\it Laboratoire de Physique Math\'ematique, 
Universit\'e Montpellier II,}}\\
{\small {\it Place E. Bataillon,  34095 Montpellier, France}}\\
{\small ${}^{d}${\it 
Laboratoire de Physique Th\'eorique, 3 rue de 
l'Universit\'e, 67084 Strasbourg, France}}}
\date{}

\maketitle
\vskip-1.0cm

\begin{abstract}
We construct new relativistic linear differential 
equation in $d$ dimensions generalizing Dirac equation
by employing the Clifford algebra 
of the cubic polynomial associated to
Klein-Gordon operator multiplied by the mass parameter.
Unlike the Dirac case where the spin content is unique  
and Lorentz covariance is manifest, 
here the spin as well as Lorentz covariance of the theory
are related to the choice of 
representation of the Clifford algebra. 
One of the considered explicit matrix representations 
gives rise to anyon-like fields in $d=1+1$.
Coupling to a U(1) gauge field is 
discussed and compared with Dirac theory.
\end{abstract}

\bibliographystyle{unsrt}    

\newpage
\section{Introduction}

A possible way to obtain relativistic wave equations is to relate
the spin degrees of freedom to an appropriate algebraic structure.
The most famous example is the Dirac equation for spin $1/2$ particles 
in which spin is associated to Clifford algebra. 
Beyond this simplest example
several possibilities have been considered, as for instance the 
Duffin-Kemmer-Petiau (DKP) algebra relevant for a first order
equation for a spin zero and spin one particles 
\cite{DKP}. These two types of algebra (Dirac and DKP) appear as special cases
of the parafermionic algebra of order $p$ ($p=1$ being the Clifford algebra
and $p=2$ the DKP one) \cite{parafermions}. 
The  parafermionic algebra is defined by cubic relations, encoding
the basic property 
that the Lorentz generators are just given by
the commutator of the generators of the algebra itself.
Besides these algebraic descriptions for integer and half-integer spin
states in any space-time dimension, specific properties appear in low
dimensions. For instance, in $2+1$ dimensions
anyons are realized on infinite dimensional representations of the universal
covering group of the $(2+1)$-dimensional Lorentz group 
\cite{anyon,spin}.
Relativistic equations for a relativistic anyon are
more involved, and various possibilities have been considered
\cite{JN,PC}. In particular, the (super)algebraic 
structure of the so called deformed Heisenberg algebra with reflection
related to parabosons and parafermions \cite{rdef}
allowed one of us to construct linear differential
equations describing universally anyons and ordinary fields of integer and
half-integer spin \cite{universal}, including topologically massive
vector gauge field \cite{topology}.

In this paper, we consider an alternative way for
obtaining a relativistic wave equation  by simple extension of
Dirac's idea.  We just write the Klein-Gordon operator as 
a perfect $n$-th power (here we study the simplest case $n=3$,
{\it i. e.} we introduce an operator $D$ such that $D^3=-m(\Box^2+m^2)$).
The present approach is based on writing the quadratic
form associated to the Minkowski metric as a cubic form. A set of
generators $g_\mu$ and $\tilde g$
associated, in turn, to the momentum vector $p_\mu$ and 
to the mass parameter $m$ 
is then introduced. This allows us to define  a first order relativistic
equation of the form $(i g^\mu \partial_\mu + m \tilde  g ) \psi=0,$
which can be treated 
as a special type of a generic Dirac-like
equations \cite{nik} 
\beq
\label{Dg}
(i \beta^\mu \partial_\mu + m \tilde  \beta ) \psi=0.
\eeq

The basic algebra (generated by $g_\mu$ and $\tilde g$)
giving rise to such a linearization
is the so called Clifford algebra of polynomials \cite{cp}. This is an $n$-th
order algebra where the generators are submitted to $n$-th order constraints.
This infinite dimensional algebra admits several finite dimensional
representations allowing us to define a first order wave equation. However,
although the algebra itself is Lorentz-invariant, the 
covariance  of the relativistic wave equation is not guaranteed for a given
representation. In other words, 
there is no guarantee to construct
Lorentz generators in terms of the $g$'s themselves. This is very 
different to the parafermionic situation. 
Various explicit types
of manifestly Lorentz covariant finite dimensional
representations of the cubic Clifford algebras 
can be obtained
and compared with the 
spinorial construction  
of the Lorentz group (related to ordinary Clifford algebras).
On the other hand, no infinite dimensional representations of
cubic Clifford algebras are known 
(even though these algebras are infinite dimensional).
In the case of a manifestly covariant equation, the approach gives us the possibility
to relate the spin degree of freedom with a new algebraic structure.
At the same time, some results concerning the cubic Clifford algebra and 
associated relativistic wave equation can be established independently
of the representation we select (or independently of the content
of the relativistic field). 

The paper is organized as follows.
In section 2, we study  the 
algebraic structure with which the Klein-Gordon
operator is written through a perfect cube. 
Some conditions are then given in order
to ensure a covariant relativistic wave equation. Two types of matrix
representations are given: one with a manifest $d$-dimensional Lorentz covariance
($d$ is arbitrary),
and another with no Lorentz covariance in $2+1$ dimensions
but admitting covariance in $1+1$ dimensions. Section 3
is devoted to the study of the relativistic equation itself.
Its solutions are given by means of appropriate projectors. 
Requirement of covariance of
the equation gives us an alternative way to calculate Lorentz
generators. 
In $1+1$ dimensions our approach gives 
a new relativistic anyon-like wave
equation for fields of spin $1/3$ and $-2/3$ 
by means of $9 \times 9$ matrices.  
We discuss also coupling  to a U(1) gauge field
and observe that new structures appear here
in comparison with Dirac theory.
Section 4 contains some conclusions and comments.

\section{Cubic root of Klein-Gordon operator}
\subsection{Clifford algebra of the polynomial}

The Clifford algebra associated with  Dirac  equation is related to
the linearization of the quadratic polynomial $P_2(x) = (x^0)^2-
(x^1)^2- \cdots -(x^{d-1})^2$. In a similar way  
there appear  the so-called Clifford algebras of  polynomials \cite{cp}.
Such algebras can naturally be defined  by introducing a series of 
generators leading to a linearization of a given polynomial of degree
$n$  with $k$ variables. From the isomorphism between degree $n$ polynomials
and symmetric tensors of order $n$, one can write any
homogeneous polynomial $P$ as
\[
P(x_1,\cdots,x_k) = \sum \limits_{\{i\}=1}^k x_{i_1} x_{i_2} \cdots x_{i_n}
g_{i_1 \cdots i_n}
\]
with $g_{i_1 \cdots i_n}$ being the  associated symmetric 
rank $n$ tensor. As for the ordinary quadratic Clifford algebra,
writing the polynomial $P$ as a perfect power of $n$ defines
the Clifford algebra ${\cal C}_P$:
\beq
\label{eq:line}
P(x_1,\cdots,x_k) = \sum \limits_{\{i\}=1}^k x_{i_1} x_{i_2} \cdots x_{i_n}
g_{i_1 \cdots i_n} = \Big(x_1 g_1 + \cdots + x_k g_k \Big)^n,
\eeq

\noi
where the generators  $g_1,\cdots, g_k$ 
are associated with the  variables $x_1,\cdots,x_k$. 
Developing explicitly the $n$-th power and identifying all the
terms leads to the relations
\beq
\label{eq:cliff}
S_n(g_{i_1}, \cdots, g_{i_n}) 
:= {1 \over n !} \sum \limits_{\sigma \in \Sigma_n}
g_{i_{\sigma(1)}} \cdots g_{i_{\sigma(n)}} = g_{i_i \cdots i_n},
\eeq
 
\noi
with $\Sigma_n$ the symmetric group of order $n$.
The Clifford algebra of the polynomial $P$ is then the order $n$ algebra
generated by the $g_i$ submitted to the constraints (\ref{eq:cliff}).
  
The algebra (\ref{eq:cliff}) is very different from the
usual Clifford algebra. Indeed, ${\cal C}_P$ is defined through $n$-th order 
constraints, and consequently the number of independent monomials
increases with polynomial's degree 
(for instance, $g_1^2 g_2$ and $g_1 g_2 g_1$ are 
independent). 
This means that we do not have  enough constraints among the 
generators to order them in some fixed way and, as a consequence,
${\cal C}_P$ turns out to be an infinite dimensional algebra.
However, it has been proved that for any polynomial a finite dimensional
(non-faithful) representation can be obtained \cite{line}. 
But, for polynomial of degree higher  than two, we do
not have  a unique representation,
and  inequivalent representations of ${\cal C}_P$
(even of the same dimension) can be
constructed (see, for instance, \cite{ineq} and below for the 
special cubic polynomials).
Furthermore, the problem of classification of the representations 
of ${\cal C}_P$ is still open, though 
it has been proved that the dimension of the
representation is a multiple of the degree of the polynomial \cite{dimrep}.
For more details one can see \cite{cliffalg} and references therein. 

\subsection{Clifford algebra of the cubic root of Klein-Gordon operator}

With the described  Clifford algebras ${ \cal C}_P$ one can represent the massive 
Klein-Gordon operator (in any  $d$-dimensional space-time) 
as a perfect $n$-th power.  
This can be achieved  with the help of
Clifford algebra of the polynomial
$P_{n,a}(p) = m^{n-2a}(p^\mu p_\mu-m^2)^{a} = 
(p_\mu g^\mu + m \tilde g)^n, 
n \ge2a$,
where $\eta_{\mu \nu} = \mathrm{diag}(1,-1,\cdots,-1)$ is
the Minkowski metric and $m$ is a mass parameter. 
Here we investigate the case $n=3$ allowing  us to take
a cubic root of Klein-Gordon operator
via the relation $m(p^\mu p_\mu-m^2)=
(p_\mu g^\mu + m \tilde g)^3$. 
In this $3$-root case, the generators $g_\mu$ and
$\tilde g$ satisfy the cubic algebra
\beq
\label{eq:cliff3}
  S_3(\tilde g,\tilde g, \tilde g) = \tilde g^3=-1, \quad
  S_3(g_\mu,\tilde g,\tilde g) = 0,\quad
  S_3(g_\mu,g_\nu,\tilde g) = {1 \over 3} \eta_{\mu \nu},\quad
  S_3(g_\mu,g_\nu,g_\lambda) = 0.  
\eeq

\noi
We denote $<g>$ the Clifford algebra generated by the $g$'s submitted
to the constraints (\ref{eq:cliff3}).
Obviously, the product

\beq
\label{eq:auto}
SO(d-1,1) \times \ZZ_2 \times \ZZ_2 \times \ZZ_3
\eeq

\noi
is an outer automorphism of this algebra. 
With respect to $SO(d-1,1)$, $g_\mu$ ($\tilde g$) are in
the vector (scalar) representation. The two $\ZZ_2$ factors 
are due to the $P$ (parity) and $T$ (time reversal) invariance of 
(\ref{eq:cliff3}). The $\ZZ_3$ automorphism is associated with
the substitution $(g_\mu,\tilde g) \longrightarrow (q g_\mu, q \tilde g)$
with $q$ a cubic root of the unity, $q^3=1$. In other words the algebra 
(\ref{eq:cliff3}) admits a $\ZZ_3$-graded structure and the $g$'s
have a gradation one.  

{}From the generators of the algebra one can construct
the  following $0$-grade vectors,
\beq
\label{eq:gamma1}
\Gamma_\mu = \tilde g^2 g_\mu, \quad
\tilde \Gamma_\mu = g_\mu \tilde g^2,\quad
\hat \Gamma_\mu = \tilde g g_\mu \tilde g. 
\eeq
Due to the algebra (\ref{eq:cliff3}), these operators
are the simplest vectors of gradation zero,
from which more complicated $0$-grade vectors can be constructed
by inserting into them $0$-grade products of three generators
$g_\lambda$, $g^\lambda$ and $\tilde g$ (with
summation in $\lambda$ implied).
The second  equation from (\ref{eq:cliff3}) means, however,
the linear dependence of the vectors (\ref{eq:gamma1}):
\beq
\label{eq:gamma2}
\Gamma_\mu + \tilde \Gamma_\mu + \hat \Gamma_\mu = 0.
\eeq

A natural question we should address is whether the automorphisms
(\ref{eq:auto}) are inner automorphisms. When we have
an inner automorphism,  this enables us to write
down the Lorentz transformations (specified by the matrix $\Lambda$) as
$\Lambda^\mu_{ \ \  \nu} g^\nu = S(\Lambda) g^\mu S^{-1}(\Lambda)$ and 
$\tilde g = S(\Lambda)\tilde g S^{-1}(\Lambda)$ with $S(\Lambda) \in <g>$.
At the infinitesimal level this is reduced to the possibility to find
the generators $J_{\mu \nu}$ such that
\beq
\label{eq:inner}
 [J_{\mu \nu}  , g_\alpha  ]  = i(\eta_{\mu \alpha} g_\nu -
\eta_{\nu \alpha} g_\mu), \quad
\big[J_{\mu \nu}, \tilde g \big] = 0, 
\eeq

\noi
with $J_{\mu \nu}$ constructed in terms of the $g$'s. 
{}From the requirement that $J_{\mu \nu}$ is a grade-zero 
antisymmetric rank 2 tensor, due to  Eq. (\ref{eq:inner})
it is constructed only with one 
$g_\mu$, one $g_\nu$ and scalars.  The only scalar we have 
are\footnote{We have also in principle the pseudo-scalar 
$\varepsilon_{{\mu_0} \cdots \mu_{\mu_{d-1}}} g^{\mu_0} \cdots g^{\mu_{d-1}}$,
with $\varepsilon_{{\mu_0} \cdots \mu_{\mu_{d-1}}}$ the totally antisymmetric 
tensor of rank $d$.} $\tilde g$ and $g^\lambda g_\lambda$.
If $J_{\mu \nu}$ does not contain the contracted product(s)
$g^\lambda g_\lambda$ (there is a strong argument in
this direction, see Eq. (\ref{eq:almost}) or Eq. (\ref{eq:j}), 
next section), then in terms of 
zero-grade vectors (\ref{eq:gamma1}) one can construct
only two possible generators commuting with 
$\tilde g$,
\beqa
\label{eq:gene}
J_{\mu \nu}^{(1)} &=&
\Gamma_\mu \Gamma_\nu + \tilde \Gamma_\mu \tilde \Gamma_\nu +\tilde \Gamma_\mu
\Gamma_\nu=\tilde g g_\mu g_\nu + g_\mu \tilde g g_\nu + g_\mu g_\nu \tilde g,
\\
J_{\mu \nu}^{(2)} &=&
\Gamma_\mu \Gamma_\nu + \tilde \Gamma_\mu \tilde \Gamma_\nu + \Gamma_\mu  
\tilde \Gamma_\nu = \tilde g^2 g_\mu \tilde g^2 g_\nu + g_\mu \tilde g^2 g_\nu
\tilde g^2 + \tilde g^2 g_\mu g_\nu \tilde g^2, \nonumber
\eeqa

\noi
where $\mu\neq\nu$ assumed.
To obtain the expression of $J_{\mu \nu}^{(1)}$ we have used (\ref{eq:cliff3}),
and it is easy to see that  $J_{\mu \nu}^{(1)}$ in 
(\ref{eq:gene}) are antisymmetric in $\mu$ and $\nu$ (see the 
second equation in (\ref{eq:cliff3})).
Using Eq. (\ref{eq:cliff3}), we get
\beq
\label{eq:almost}
\big [J_{\mu \nu}^{(1)}, g_\alpha \big ] =
2(\eta_{\mu \alpha} g_\nu - \eta_{\nu \alpha} g_\mu) -
[g_\mu g_\nu g_\alpha + g_\mu g_\alpha g_\nu + 
g_\alpha g_\mu g_\nu , \tilde g],\ \ \
\big [J_{\mu \nu}^{(1)}, \tilde g \big ] =0. \nonumber
\eeq

\noi
One could directly calculate $[J_{\mu \nu}^{(2)}, g_\alpha]$ 
to find that $J_{\mu \nu}^{(2)}$
does not fulfill either the correct commutation relations with  $g_\alpha$
(we already know that $[J_{\mu \nu}^{(2)},\tilde  g ] =0$).
Instead, it is more easy to check this
on explicit example 
(see the $9 \times 9$ matrix representation
hereafter (\ref{eq:9d})).
So, neither $J_{\mu \nu}^{(1)}$ nor 
$\Big(J_{\mu \nu}^{(2)}-J_{\nu \mu}^{(2)}\Big) $ 
(the antisymmetrized version of $J_{\mu \nu}^{(2)}$) could be the Lorentz
generators.
However, if the second part of the commutator of $J_{\mu \nu}^{(1)}$ with 
$g_\alpha$ in (\ref{eq:almost}) vanishes, then  $J_{\mu \nu}^{(1)}$ 
are the Lorentz
generators (as it is the case for the representation (\ref{eq:dirac}), see
below).
Conversely, if we assume that one can find $J_{\mu \nu}$ satisfying 
(\ref{eq:inner}), it is tempting to conclude that this implies 
vanishing the  second part of the commutator of $J_{\mu \nu}^{(1)}$ with 
$g_\alpha$ (see Eq. (\ref{eq:almost})). Indeed, if  we introduce 
$J_{\mu \nu}$ as  a $0$-grade operator  constructed with one 
$g_\mu$, one $g_\nu$ and the scalars $\tilde g$ and $g^\lambda g_\lambda$, 
from the Lorentz covariance we find that $J_{\mu \nu}^{(1)}$
transforms as a rank 2 tensor. So, 
$[J_{\mu \nu}, J_{\alpha \beta}^{(1)}]$ has a simple form.
But now, if using (\ref{eq:almost}) we calculate 
$[J^{(1)}_{\mu \nu}, J_{\alpha \beta}]$,
this gives a very complicated expression 
(coming from  $[J^{(1)}_{\mu \nu}, g^\lambda \ldots g_\lambda]$), 
and the cancellation 
seems to be difficult to obtain except if 
$[g_\mu g_\nu g_\alpha + g_\mu g_\alpha g_\nu + g_\alpha g_\mu g_\nu , 
\tilde g] =0$.  In such a case
$J_{\mu \nu}$ is equal to ${i \over 2} J_{\mu \nu}^{(1)}$.
Furthermore, the results of section 2 also suggest that $J_{\mu \nu}$
does not contain $g^\lambda \ldots g_\lambda$.

Consequently, we could conjecture that the outer 
automorphism $SO(d-1,1)$ becomes an inner automorphism if and only if
$[g_\mu g_\nu g_\alpha + g_\mu g_\alpha g_\nu + g_\alpha g_\mu g_\nu , 
\tilde g] =0$, and then we have

\beq
\label{eq:generators}
J_{\mu \nu} = {i \over 4} \Big(\tilde g g_{[\mu} g_{\nu ]} + 
g_{[ \mu} \tilde g g_{\nu ]} + g_{[ \mu} g_{\nu ]} \tilde g \Big),
\eeq

\noi
with $g_{[\mu} g_{\nu ]}= g_\mu g_\nu - g_\nu g_\mu$.

So, on the one hand, for a given representation ${\cal R}_g$ of $<g>$,
$g_\mu, \tilde g \longrightarrow G_\mu, \tilde G$, there is no guarantee 
to find
the generators $J_{\mu \nu}$ constructed from the $g$'s. In such a case
the representations  ${\cal R}_g = <G_\mu, \tilde G>$ and 
${\cal R}_g^\Lambda= <\Lambda^\mu_{ \ \  \nu} G^\nu, \tilde G>$ 
are inequivalent. This is
very different to the usual Clifford algebra case\footnote{However, when the 
space-time 
dimension is odd, we have two inequivalent representations (equivalent to
$\pm \gamma_\mu$ with $\gamma_\mu$ the Dirac matrices) 
because there is no $\gamma_5$-like matrix (chirality).
Thus, the  $T$ inversion is an outer automorphism
for the usual Clifford algebra in odd dimensions.}.
But, on the other hand, if $SO(d-1,1)$ is an inner automorphism, this means that
${\cal R}_g = <G_\mu, \tilde G>$ and 
${\cal R}_g^\Lambda= <\Lambda^\mu_{ \ \  \nu} G^\nu, \tilde G>$ are equivalent
representations, or that the $G$'s act on an appropriate representation
of the Lorentz group. This constitute an adapted extension of the spinorial
construction of the Lorentz group (related to the cubic instead of 
quadratic Clifford algebras). 

\subsection{Examples of $<g>$-representations}

As we noted above, representations of the Clifford algebras of polynomials
are not classified and only some special matrix representation are known. 
For $<g>$ we have found two types of representations.
The first one is constructed with the usual Dirac matrices in any space-time
dimension,
\beqa
\label{eq:dirac}
\begin{array}{ll}
G_\mu = \pmatrix{0&\gamma_\mu&0 \cr
                 0&0& \gamma_\mu \cr
                 0&0&0}, & 
\tilde G = \pmatrix{0&i \gamma_{d+1}&0 \cr
                 0&0&i  \gamma_{d+1} \cr
                 1&0&0}. 
\end{array}
\eeqa 

\noi
Because of the presence of the
matrix $\gamma_{d+1}$,  in odd-dimensional space-time
we assume that $\gamma_\mu$ 
are realized via the gamma-matrices $\gamma^{(d-1)}_\mu$ 
corresponding to dimension $(d-1)$, e.g., in the form
$\gamma_\mu=\gamma^{(d-1)}_\mu\otimes \sigma_1$,
whereas $\gamma_{d+1}$ can be chosen as $\gamma_{d+1}=1\otimes \sigma_3$.
It is easy to see that in this Dirac-like representation 
both $J_{\mu \nu}^{(1)}$ and $J_{\mu \nu}^{(2)}$ in 
Eq. (\ref{eq:gene}) give
$J_{\mu \nu} =- {i \over 2} \gamma_\mu \gamma_\nu \otimes  
{\mathrm diag}(1,1,1).$

The second representation is obtained by linearizing firstly 
the polynomial $m((p_0)^2 - (p_1)^2 - \cdots )$, 
\[
m((p_0)^2 - (p_1)^2 -
\cdots )=
\pmatrix{0&m&0\cr0&0&p_0+p_1\cr p_0-p_1&0&0}^3 +\cdots 
\]
with subsequent linearization of this sum of perfect cubes 
by means of the twisted tensorial product \cite{ineq}. 
As a result, we end up with the 
$3^{[{d+2 \over 2}]} \times 3^{[{d+2 \over 2}]}$-dimensional matrices,
with $[a]$ being the integer part of $a$.
In $2+1$ dimensions, the corresponding representation
is given by the $9 \times 9$ matrices:
\beqa
\label{eq:9d}
\begin{array}{ll}
G_\rho = \pmatrix{ 
T_\rho&{\bf 0}&{\bf 0}\cr
{\bf 0}&qT_\rho&{\bf 0}\cr
{\bf 0}&{\bf 0}&q^2T_\rho},\, \, 
\rho=0,1,\, \,
G_2 = \pmatrix{ 
{\bf 0}&{\bf 0}&{\bf 0}\cr
{\bf 0}&{\bf 0}&I\cr
I&{\bf 0}&{\bf 0}
},\, \,
\tilde G = 
\pmatrix{{\tilde T}&-I&{\bf 0}\cr
{\bf 0}&q\tilde T&iI\cr
-iI&{\bf 0}&q^2\tilde T
},
\end{array}
\eeqa

\noi
where $q$ is a primitive cubic root of the unity,
${\bf 0}$ and $I$ are zero and unit $3\times 3$ matrices,
and 

\beqa
\label{eq:9d+}
\begin{array}{ll}
T_0 = \pmatrix{ 
0&0&0\cr
0&0&1\cr
1&0&0
},\quad
T_1 = \pmatrix{ 
0&0&0\cr
0&0&-1\cr
1&0&0
},\quad
\tilde T = \pmatrix{ 
0&1&0\cr
0&0&0\cr
0&0&0}.
\end{array}
\eeqa

\noi
On this explicit representation, one can check that $J_{\mu \nu}^{(1)}$
and 
$\Big(J_{\mu \nu}^{(2)}-J_{\nu \mu}^{(2)}\Big)$
defined by (\ref{eq:gene}) do not satisfy the correct 
commutation relation  with $G_\alpha$. Moreover, one can also directly
check that there does not exist any $9\times 9$ matrix $J_{\mu \nu}$ satisfying
(\ref{eq:inner}). Therefore, for this representation $SO(2,1)$ is an outer
automorphism.

Representation (\ref{eq:9d}) turns out to be more interesting
in the case of $1+1$ dimensions. 
Indeed, the matrices
(\ref{eq:9d}) lead to different types of $9 \times 9$ dimensional 
representations. The first type is given by  the set of generators
${\cal R}_{02} =
<G_0,G_2,\tilde G>$ or by the set
${\cal R}_{12}= <e^{i \pi/3} G_1,G_2,\tilde G>$. 
For both these representations,
there is no Lorentz generator satisfying Eq. (\ref{eq:inner}).
The second type representation is characterized by the
set 
\beq
\label{R01}
{\cal R}_{01} = <G_0,G_1,\tilde G>,
\eeq
and by the Lorentz generator
\beq
\label{spin1/3}
J_{01} = \frac{i}{3}\pmatrix{1&0&0 \cr 0&1&0 \cr
0&0&-2} \otimes \pmatrix{1&0&0 \cr 0&1&0 \cr 0&0&1},
\eeq
which satisfies
the correct commutation relations with $G_0, G_1, \tilde G$,
i.e. for this representation the automorphism $SO(1,1)$ is internal.
Traceless generator (\ref{spin1/3}) is obtained
via Eq. (\ref{eq:generators})
with appropriate `renormalization' consisting in a  
shift for the matrix proportional
to the unit one.

Finally, let us note that in $0+1$ dimension,
there exists the $3 \times 3$ representation of $<g>$ given by
the matrices

\beq
\begin{array}{ll}
G_0 = \pmatrix{0&1&0 \cr 0&0&1 \cr 0&0&0},\quad &
\tilde G =  \pmatrix{0&1&0 \cr 0&0&-1 \cr 1&0&0}.
\end{array}
\eeq

To conclude the discussion of the representations 
of the Clifford algebra $<g>$,
one notes that
only (some) finite dimensional representations of Clifford algebras 
have been found.
On the other hand, as we noted above, 
in $2+1$ dimensions anyons are related to 
infinite dimensional representation of 
$\overline{SL(2,\RR)}$ (the universal covering
of Lorentz group) \cite{anyon,spin}. 
So, it would be very interesting to try to find infinite  
dimensional representations of $<g>$ in relation to the 
infinite dimensional 
representation of $\overline{SL(2,\RR)}$.

\section{Cubic root of Klein-Gordon equation}
\subsection{First order equations and almost Dirac algebras}
After the formal discussion of the algebra $<g>$ and its representations, 
now we are ready to define
an adapted extension of the Dirac operator being a cubic root, instead
of a quadratic one, of the Klein-Gordon operator,
\beq
\label{eq:cubic}
D=(i g^\mu \partial_\mu + m \tilde g),\quad
D^3= -m(\Box^2+m^2), 
\eeq

\noi
with the $g$'s defined in the previous section.
This operator enables us to define two possible equations 
generalizing the Dirac one and its conjugate,
\beq
\label{eq:trirac}
(i g^\mu \partial_\mu + m \tilde g) \psi =0, \quad
\bar \psi (i \overleftarrow{\partial_\mu} g^\mu+ m \tilde g) = 0,
\eeq

\noi
where $\bar \psi \ \overleftarrow{\partial_\mu} \equiv \partial_\mu \bar \psi$.
Following the general discussion of the previous section, if for a given
representation of $<g>$ the outer automorphism $SO(d-1,1)$ is promoted to
an inner one, then the fields $\psi$ and $\bar \psi$ turn out to be in
some representation ${\cal D}$ and ${\cal D}^\star$ (the dual representation
of ${\cal D}$) of $\overline{SO(d-1,1)}$.
In other words, under a Lorentz transformation $\Lambda$ the fields
transform as $S(\Lambda) \psi$ and $\bar \psi S^{-1}(\Lambda)$. This means that
the equations (\ref{eq:trirac}) constitute  alternative relativistic
equations. In the opposite case, (\ref{eq:trirac}) has no Lorentz covariance.
Some general results upon this equation can be established without
specifying concrete representation for $<g>$.
Similar equations already appeared in \cite{1dfsusy} in the world-line
formalism  (for the massless case)  in the
context of fractional supersymmetry \cite{fsusy}.

If we multiply the first equation by $\tilde g^2$ on the left and the second
on the right, we get 
\beq
\label{eq:trirac2}
(i \Gamma^\mu \partial_\mu - m) \psi = 0, \quad
\bar \psi (i \overleftarrow{\partial_\mu} \tilde \Gamma^\mu- m)   = 0,
\eeq

\noi
with the matrices $\Gamma_\mu$ and $\tilde \Gamma_\mu$
defined in (\ref{eq:gamma1}). A direct calculation
with  the algebra (\ref{eq:cliff3}) 
leads to the almost Dirac algebras
\beq
\label{eq:almostdirac}
\{\Gamma_\mu,\Gamma_\nu\} = 2 \eta_{\mu \nu} -\{g_\mu,g_\nu\} \tilde g,\quad
\{\tilde \Gamma_\mu,\tilde \Gamma_\nu\} = 2 \eta_{\mu \nu} -
\tilde g\{g_\mu,g_\nu\}.
\eeq

\noi
Introducing the time-like  momentum $p^\mu$  $(p_\mu p^\mu=m^2,$ $m \ne 0)$,
from 
(\ref{eq:almostdirac}) 
we get 
$\Big(p^\mu \Gamma_\mu\Big)^2=m^2\left(1-({e^\mu} 
g_\mu)^2 \tilde g\right)$
and $\Big(p^\mu \tilde \Gamma_\mu\Big)^2=m^2\left(1-\tilde 
g({e^\mu} g_\mu)^2\right)$ with $e^\mu \equiv p^\mu/m$. 
Below we shall shaw (see Eq. (\ref{eq:proj}))
that the operators 
$\left(1-({e^\mu} g_\mu)^2 
\tilde g \right)$ 
and $\left(1-\tilde g({e^\mu} g_\mu)^2\right)$ 
are the projectors 
onto the physical space for $\psi$ and $\bar \psi$, respectively. 
We call $\Gamma_\mu$ and 
$\tilde\Gamma_\mu$ the left and right almost Dirac matrices since
being restricted to the 
physical subspaces, they satisfy a usual quadratic Clifford algebras.

\subsection{Projectors}

To solve  Eq. (\ref{eq:trirac}), we introduce two
series of projectors (for $\psi$ and $\bar \psi$),

\beqa
\label{eq:proj}
\begin{array}{llll}
\Pi_\pm(p) &= -{1 \over 2}\Big(\pm {e^\mu} g_\mu
+\tilde g\Big)^2\tilde g&
\Pi^\star_\pm(p)&= -{1 \over 2}\tilde g\Big(\pm {e^\mu} g_\mu +
\tilde g\Big)^2 
\cr
 &={1 \over 2} \left((1-\left( {e^\mu} g_\mu \right)^2 \tilde g \pm
\tilde g^2\left( {e^\mu} g_\mu \right) \right),&
&={1 \over 2} \left((1-\tilde g\left( {e^\mu} g_\mu \right)^2  \pm
\left( {e^\mu} g_\mu \right)\tilde g^2 \right), \cr
\Pi_0(p)&=\left( {e^\mu} g_\mu \right)^2 \tilde g,&
\Pi_0^\star(p)&=\tilde g\left( {e^\mu} g_\mu \right)^2. \cr
\end{array}
\eeqa

\noi
Using the outer automorphism $SO(d-1,1)$, we can consider 
$\Pi_\varepsilon(\hat p)$ and 
$\Pi^\star_\varepsilon(\hat p)$, $\varepsilon=\pm,0$,
with $\hat p=(m,0,\cdots,0)$ the rest frame momentum. Then, it is easy to prove
that
$\Pi_\varepsilon(\hat p) \Pi_{\varepsilon^\prime}(\hat p) = 
\delta_{\varepsilon 
\varepsilon^\prime} \Pi_\varepsilon(\hat p)$ and  
$\Pi_+(\hat p)+\Pi_-(\hat p)+
\Pi_0(\hat p)=1$ 
(and similar relations for $\Pi^\star(\hat p)$), i.e.
$\Pi_\varepsilon(p)$ and  $\Pi^\star_\varepsilon(p)$ 
constitute two complete sets of projectors. 
Denoting, for a
given $k$-dimensional representation, 
$n_\epsilon = {\mathrm Tr}
(\Pi_\epsilon(\hat p)) = 
{\mathrm Tr}(\Pi_\epsilon( p))$, we have obviously $n_+ + n_- +n_0 = k$ and 
$n_+=n_-$.
Moreover, using (\ref{eq:cliff3}) we have $n_0 = k/3$ (this is
enough to prove that
$k$ is a multiple of $3$), and finally we get
\beq
\label{eq:trace}
{\mathrm Tr}(\Pi_0(p))= {\mathrm Tr}(\Pi_+(p))={\mathrm Tr}(\Pi_-(p))=n=\frac{1}{3}k,
\eeq

\noi
and similarly for $\Pi^\star$. 

\subsection{Solutions to the equation (\ref{eq:trirac})}

The projectors $\Pi$ and $\Pi^\star$
are useful to calculate the solutions to Eq. (\ref{eq:trirac}).
{}From now on we just consider the equation for $\psi$, the other one
is  totally similar. 
Let us introduce the solutions
of (\ref{eq:trirac}) in the form of the plane waves of positive 
and negative energy, 
$\psi_\pm(p) = e^{\mp i p^\mu x_\mu} W_\pm(p)$. 
Then, the equation is reduced to the equations 
for $W_\pm(p)$,
$
(\pm  p^\mu  g_\mu +m \tilde g) W_\pm(p)=0.
$
{}From the relation 
$
(\pm  p^\mu  g_\mu +m \tilde g)^3=m(p^\mu p_\mu-m^2)=0,
$
it is easy to see that 
$
W_\pm(p) \in {\mathrm Im} \Pi_\pm(p).
$
Finally, from $(p^\mu g_\mu)^3=0$ we observe that the set of vectors
$W_0(p) \in {\mathrm Im} \Pi_0(p)$ are neither positive energy solutions
nor negative energy ones, but the auxiliary fields. There the equations of
motion are reduced to
$
W_0(p)=0.
$
Since, $\Pi_\varepsilon(p)$ constitute a complete set of projectors, the
space of solutions decomposes into $S= {\mathrm Im} \Pi_+(p) \oplus
{\mathrm Im} \Pi_-(p)\oplus {\mathrm Im} \Pi_0(p)$.
The projectors $\Pi_\pm(p)$ are related to the physical solutions,
{i. e.} to the positive and negative energy solutions,
whereas the
projector $\Pi_0(p)$ characterizes the auxiliary fields.
Differently to the quadratic Clifford algebra, 
here we have auxiliary fields 
(like they are present in generic Dirac-like equation (\ref{Dg})).
This comes from the fact that the $g_\mu$ are singular, $(g_\mu)^3=0$.
This singularity, in turn, is due to the fact that basically we have a quadratic
form (related to the Minkowski norm) which we write through a `cubic'
form.

For the $d$-dimensional representation (\ref{eq:dirac})
the equation (\ref{eq:trirac}) takes a simple
form. Writing $\psi^t=\pmatrix{\psi_1 & \psi_2 & \psi_3}$,
it gives
\beq
\label{eq:dirac-eq}
\psi_1= 0, \qquad
(i \gamma_\mu \partial_\mu + m \gamma_{d+1}) \psi_i =0, \quad i=2,3.
\eeq

\noi
Hence, $\psi_1$ are  auxiliary fermionic fields and $\psi_2,\psi_3$
are the usual Dirac fields. 

For the representation (\ref{eq:9d}), we have a relativistic wave equation
in $2+1$ dimensions with Lorentz invariance only in the $(x_0,x_1)$
subspace.
For the representation 
(\ref{R01}) in $1+1$ dimensions with
a basis where $\Pi_0(\hat p)={\mathrm diag}
{\mathrm} (1,1,1,0,0,0,0,0,0)$, 
the Lorentz generator takes a simple
form
$$J=\frac{i}{3}{\mathrm diag}(1,1,1;1,-2,1;-2,1,-2).$$

\noi 
Therefore, it seems that  the physical fields 
can be treated as $(1+1)$-dimensional anyons  \cite{grz}
of spin $1/3$, $-2/3$. We shall return to this point in
last section.

\subsection{Alternative calculation of Lorentz generators}

Assuming that for a given representation of $<g>$ the Lorentz
covariance of Eq. (\ref{eq:trirac}) is manifest, we have an alternative
way to calculate the Lorentz generators. Indeed, we solve the equation
(\ref{eq:trirac}) in any frame using the projectors $\Pi(p)$. 
Let us choose 
two particular frames: the rest frame 
and a frame such that $p^\mu=(
p_0 \sim m, p_1 \ll m,0,\cdots,0)$. We denote respectively 
$W_\varepsilon(\hat p)$ and $W_\varepsilon(p)$ the corresponding solutions,
such that $S_\varepsilon(\hat p) = \Big\{ W_\varepsilon^i(\hat p), 
i=1,\cdots,n \Big\}$ and $S_\varepsilon( p) = \Big\{ W_\varepsilon^i( p), 
i=1,\cdots,n \Big\}$ constitute a complete basis of solutions in these 
two frames. 
Now we have two ways to write $W_\varepsilon(p)$. On the one hand,
we have
\beq
\label{eq:sol0-p1}
W_\varepsilon(p) = C_{\varepsilon,  p} \Pi_\varepsilon(p) 
W_\varepsilon(\hat p),
\eeq

\noi
with $C_{\varepsilon, p}$ a constant of normalization ($C_{\varepsilon, p}
\sim 1,$ for $p \sim \hat p$).  
On the other hand,
if we denote $B(p)$ the boost from the rest
frame to the moving frame,  we have
\beq
\label{eq:sol0-p2}
W_\varepsilon(p) = B(p) W_\varepsilon(\hat p).
\eeq

\noindent
Comparing Eqs. (\ref{eq:sol0-p1}) and (\ref{eq:sol0-p2}),
we obtain
\beq
\label{eq:boost}
B(p) = \Pi_+(p) \Pi_+(\hat p) +  \Pi_-(p) \Pi_-(\hat p) +  
\Pi_0(p) \Pi_0(\hat p).
\eeq

\noi
For the infinitesimal boost along the $x_1$ axis,
$\hat p^\mu = (m,0,\cdots,0) \longrightarrow p^\mu = (E \sim m ,p \sim 0,
0,\cdots,0)$,
developing (\ref{eq:boost}) at the first order gives
\beq
\label{eq:boost1}
B(p) = 1 -i {p \over m} J_{01}.
\eeq

\noi
If we proceed along these lines for an arbitrary solution, {\it i. e.}
with or without a manifest Lorentz covariance, the expression of $J_{01}$
(and in a similar way of $J_{0i}$) looks not so symmetric
(but at that point the algebra (\ref{eq:cliff3}) is not
used to simplify its expression). However, for the
two solutions where the Lorentz covariance is guaranteed 
(like for representation
(\ref{eq:dirac}) and
${\cal R}_{01}$ in $1+1$ dimensions),
$J_{01}$ is reduced to a very simple expression
(all other terms vanish):

\beq
\label{eq:j}
J_{01}^p = -{i \over 2} \left(\tilde g g_1 g_0 + g_1 \tilde g g_0 \right).
\eeq

\noi
Compared with $J_{01}$ in (\ref{eq:generators}),
the term $ -{i \over 2}g_1 g_0  \tilde g$ is 
missing. Obtaining $J_{01}^p$, 
we started from
the equation (\ref{eq:trirac}) without any reference to the algebra
(\ref{eq:cliff3}). {}From the equation (\ref{eq:trirac}), if we do not make
any reference to (\ref{eq:cliff3}),  only 
$\tilde g^2 g_\mu$ (or $g_\mu \tilde g^2$) is a vector, but from the algebra 
(\ref{eq:cliff3}) $g_\mu$ is a vector and $\tilde g$ is a scalar. 
It can be checked explicitly that $J_{\mu \nu}^p$ in (\ref{eq:j}) and
$\tilde g^2 g_\mu$ satisfy the correct commutation relations,
although $g_\mu$ and $\tilde g$ with $J_{\mu \nu}^p$ do not.
So, the generators $J_{\mu \nu}^p$ obtained  in (\ref{eq:j}) can be called the 
{\it partial} generators and $J_{\mu \nu}$ in (\ref{eq:generators})  
the {\it full} ones. 
Even if we cannot obtain the {\it full} generators through this
process, it gives some useful information. Indeed, in the previous section 
we have raised the possibility that $J_{\mu \nu}$ could (in principle)
contain terms like $g_\lambda g^\lambda$. But such terms never appear with
(\ref{eq:boost}), and this supports the conjecture
(on the form of $J_{\mu \nu}$) of the previous section.

\subsection{U(1) interaction}
Let us consider the problem of coupling the field system
described by the cubic root equation (\ref{eq:trirac}) to the external
U(1) gauge field $A_\mu$. Introducing the covariant derivative 
$\nabla_\mu=\partial_\mu - ie A_\mu$ 
($e$ being the charge of the field $\psi$), 
the equation becomes
\beq
\label{eq:u(1)}
(i g^\mu \nabla_\mu + m \tilde g) \psi =0.
\eeq

\noi
The  calculation of $(i g^\mu \nabla_\mu + m \tilde g)^3$
gives
\beq
\label{eq:u(1)2}
(i g^\mu \nabla_\mu + m \tilde g)^3=-i(g_\mu \nabla^\mu)^3
-m(\nabla_\mu \nabla^\mu +m^2  
- e   F^{\mu \nu} J_{\mu \nu}),
\eeq

\noi
with $F_{\mu \nu} = \partial_\mu A_\nu - \partial_\nu A_\mu$, and 
$J_{\mu \nu}$ given by (\ref{eq:generators}). 
Due to the algebra of generators (\ref{eq:cliff3}),
the first term from Eq. (\ref{eq:u(1)2}) is reduced to
\beq
\label{theta}
-i(g_\mu \nabla^\mu)^3=-\frac{1}{4}e (F^{\mu\nu}\nabla^\lambda
(g_{[\mu} g_{\nu]} g_\lambda
+g_{[\mu}g_\lambda g_{\nu]}+
g_\lambda g_{[\mu} g_{\nu]})
-e{\cal T}(A),\, \,
{\cal T}=(\partial^\mu\partial^\nu A^\lambda)g_\mu g_\nu g_\lambda.
\eeq
The last term in Eq. (\ref{theta}) seems to be not
manifestly gauge-invariant, but its change under 
the gauge transformation
$A_\mu\rightarrow A_\mu+\partial_\mu\Lambda$ disappears 
due to the algebra (\ref{eq:cliff3}): 
$\partial_\mu\partial_\nu\partial_\lambda\Lambda g^\mu g^\nu g^\lambda=0$.
With this observation   
the term ${\cal T}$ can be transformed 
identically due to the algebra (\ref{eq:cliff3}) 
into the manifestly gauge-invariant form ${\cal T}=\frac{1}{6}
\partial_\lambda F_{\mu\nu}(g^\lambda g^{[\mu}g^{\nu]}+
g^{[\mu}g^\lambda g^{\nu]})$.
Summarizing all this, we conclude that 
the modified Klein-Gordon equation corresponding to 
our field system interacting with  the U(1) gauge field
via the prescription of minimal coupling (\ref{eq:u(1)})
is given by
\beqa
\label{kgu1}
\Big(
\nabla_\mu\nabla^\mu +m^2
&-&eF^{\mu\nu}J_{\mu\nu} 
\nonumber\\
&+&\frac{e}{2m}F^{\mu\nu}
(g_\lambda g_{\mu} g_{\nu}+
g_{\mu}g_\lambda g_{\nu}+g_{\mu} g_{\nu} g_\lambda)
\nabla^\lambda 
\nonumber\\
&+&\frac{e}{3m}\partial^\lambda F^{\mu\nu}(g_\lambda g_{\mu}g_{\nu}+
g_{\mu}g_\lambda g_{\nu})\Big)\psi=0.
\eeqa
Therefore, besides the 
standard spin-field coupling described by the third term
in modified Klein-Gordon equation
(which appears in the case of 
Dirac field with minimal coupling prescription),
here in generic case we have new structures
given by the 4-th and 5-th terms in Eq. (\ref{kgu1}).
But direct checking shows that in  
representations (\ref{eq:dirac})
and (\ref{R01}), for which Lorentz symmetry is manifest, 
the relation $g_\mu g_\nu  g_\lambda =0$ is valid.
Therefore, in these cases the modified
Klein-Gordon equation has the same structure
as the quadratic equation  corresponding to
Dirac equation with minimal coupling.
This means, in particular, that  the corresponding 
Lorentz covariant field systems  
are characterized by the gyromagnetic ratio
$g=2$. 

In the peculiar (1+1)-dimensional case (\ref{R01}), because 
of the `renormalization' of 
(\ref{eq:generators}), the equation takes the form

\beq
\label{kgu2}
\Big(
\nabla_\mu\nabla^\mu +m^2
-eF^{01}J_{01} -\frac {e}{3} F^{01}\Big)\psi=0,
\eeq

\noi
with $J_{01}$ given by (\ref{spin1/3}).

\section{Concluding remarks}

Clifford and Grassmann algebras are the basic ingredient for the spinorial
representation of the Lorentz group and  they naturally appear as
fundamental generators for the Dirac operator. 
In this paper, we have constructed 
a new wave equation in line of Dirac equation.
Indeed, from Clifford algebra of a polynomial of degree higher than two
(cubic in our case) we were able to define a relativistic wave equation 
involving the cubic root of the Klein-Gordon operator instead of a
square root. Moreover, as Dirac equation is related to 
supersymmetry,
it has been observed that the equation considered in this paper can
be related to an extension of supersymmetry involving an $n$-th
order algebra, namely fractional supersymmetry \cite{1dfsusy, fsusy}.
Conversely, it is well known that supersymmetry is related to
Clifford and Grassmann algebra. Similarly it has also been established
that fractional supersymmetry is connected to Clifford algebras
of polynomials \cite{cliffalg}. So, all these $n$ order structures
(Clifford algebra of polynomials,  fractional supersymmetry and $n$-th
root of the Klein-Gordon operator) are interconnected. 

On a more practical ground,  within the framework of the Clifford
algebra of the polynomial $m(p_\mu p^\mu -m^2)$ the relativistic
equation generalizing Dirac equation  has been obtained. Differently to
the quadratic case, where the spin content is unique (spin-$1/2$ particles),
here the spin content is related to the representation of the Clifford
algebra we take. Only finite dimensional representations are known
for this type of algebras.  Two explicit matrix representations
are given here, and one of them gives an appropriate equation for
$(1+1)$-dimensional anyons of spin $1/3$  and $-2/3$. 

Constructing the corresponding Lorentz generator for the 
$1+1$ anyonic case,
we fixed the corresponding matrix to be traceless.
So, strictly speaking, the question on spin content
of such anyonic-like fields deserves further investigation.
This point can be clarified by realizing
secondary quantization to reveal
the corresponding spin-statistics relation. 
Since the massive field theory in $1+1$ dimensions 
can be treated as a reduction of the corresponding massless theory
in $2+1$ dimensions, here the analysis of ref. \cite{dj}
on statistics for massless $(2+1)$-dimensional theories 
may be very helpful.

No infinite dimensional representation of the 
infinite Clifford algebra $<g>$
has been found here. It would be very
interesting to construct such representations in the context of
$(2+1)$-dimensional anyons.

Finally, we could have chosen a slightly different cubic Clifford algebra,
for which the massive and massless cases would have been put on the 
same footing. Introducing some universal mass parameter $M$
(it could be, e.g., the Planck mass), one can consider 
the Clifford algebra
$M(p_\mu p^\mu-m^2)=(p_\mu g^\mu + m \tilde g + M \hat g)^3$.


\paragraph*{Acknowledgements}
Ph. Revoy is gratefully acknowledged for useful discussions.
One of us (MRT) would like to thank USACH for its hospitality, where
the part of this work was realized. The
work was supported in part  by 
the grants 1980619 and 7980044 from FONDECYT (Chile)
and by DICYT (USACH).


\end{document}